%% file: main.tex
\documentclass{article}

    \PassOptionsToPackage{numbers, compress}{natbib}

\usepackage[final]{neurips_2024}
\usepackage{graphicx}
\usepackage{amsmath}
\usepackage{float}
\usepackage{subcaption}
\usepackage{enumitem}
\usepackage{multirow}

\usepackage[utf8]{inputenc} 
\usepackage[T1]{fontenc}    
\usepackage{hyperref}       
\usepackage{url}            
\usepackage{booktabs}       
\usepackage{amsfonts}       
\usepackage{nicefrac}       
\usepackage{microtype}      
\usepackage{xcolor}         
\usepackage[ruled,vlined]{algorithm2e}
\usepackage{tcolorbox}

\title{Defending Large Language Models Against Jailbreak Exploits with Responsible AI Considerations}

\author{
  Wong Ryan \quad
  Hosea David Ng Yu Fei \quad
  Dhananjai Sharma \\
  \textbf{Glenn Ng Jun Jie} \quad
  \textbf{Kavishvaran Srinivasan} \\
  National University of Singapore \\
  \texttt{\{e1374031, e0774652, e1373809, e0969111, e1373770\}@u.nus.edu}
}

\begin{document}
\maketitle

\begin{abstract}
Large Language Models (LLMs) remain susceptible to jailbreak exploits that bypass safety filters and induce harmful or unethical behavior. This work presents a systematic taxonomy of existing jailbreak defenses across prompt-level, model-level, and training-time interventions, followed by three proposed defense strategies. First, a \emph{Prompt-Level Defense Framework} detects and neutralizes adversarial inputs through sanitization, paraphrasing, and adaptive system guarding. Second, a \emph{Logit-Based Steering Defense} reinforces refusal behavior through inference-time vector steering in safety-sensitive layers. Third, a \emph{Domain-Specific Agent Defense} employs the MetaGPT framework to enforce structured, role-based collaboration and domain adherence. Experiments on benchmark datasets show substantial reductions in attack success rate, achieving full mitigation under the agent-based defense. Overall, this study highlights how jailbreaks pose a significant security threat to LLMs and identifies key intervention points for prevention, while noting that defense strategies often involve trade-offs between safety, performance, and scalability. Code is available at: \url{https://github.com/Kuro0911/CS5446-Project}
\end{abstract}

\input{main/Introduction}

\input{main/Background}
\input{main/Proposed}

\input{main/Results}
\input{main/Conclusion}

\bibliographystyle{vancouver}
\bibliography{ref}
\input{Appendix}
\end{document}

%% file: main/Introduction.tex
\section{Introduction}


Large language models (LLMs) have demonstrated remarkable capabilities in various tasks including natural language understanding, generation, and reasoning, making them increasingly integral to applications ranging from conversational agents to decision support systems. However, the widespread deployment of LLMs also exposes them to adversarial manipulation, including \emph{jailbreak} exploits that are carefully crafted prompts that bypass safety filters and induce the model to produce unsafe, biased, or otherwise harmful outputs. Such vulnerabilities pose significant risks to privacy, ethics, and user trust.

Addressing these challenges requires a proactive and systematic approach that integrates \emph{Responsible AI} principles, encompassing fairness, transparency, accountability, and robustness. Traditional post-hoc interventions, such as output moderation or input sanitization, are often reactive and insufficient against sophisticated jailbreak attacks that exploit the model's internal behaviors. Consequently, there is a growing need to explore defenses that are embedded directly into the model’s training, architecture, and optimization processes.

In this project, we first establish a comprehensive taxonomy of jailbreak methods and training-time defenses, followed by a proposal of robust interventions  designed to enhance safety and alignment. We then empirically evaluate these methods through controlled experiments to measure their effectiveness against representative jailbreak exploits. Finally, we draw insights and recommendations for deploying LLMs in alignment with Responsible AI principles, aiming to foster both robustness and trustworthiness in real-world applications.

%% file: main/Background.tex
\section{Taxonomy of Jailbreak Defense}

\begin{figure}[H]
    \centering
    \includegraphics[width=\linewidth]{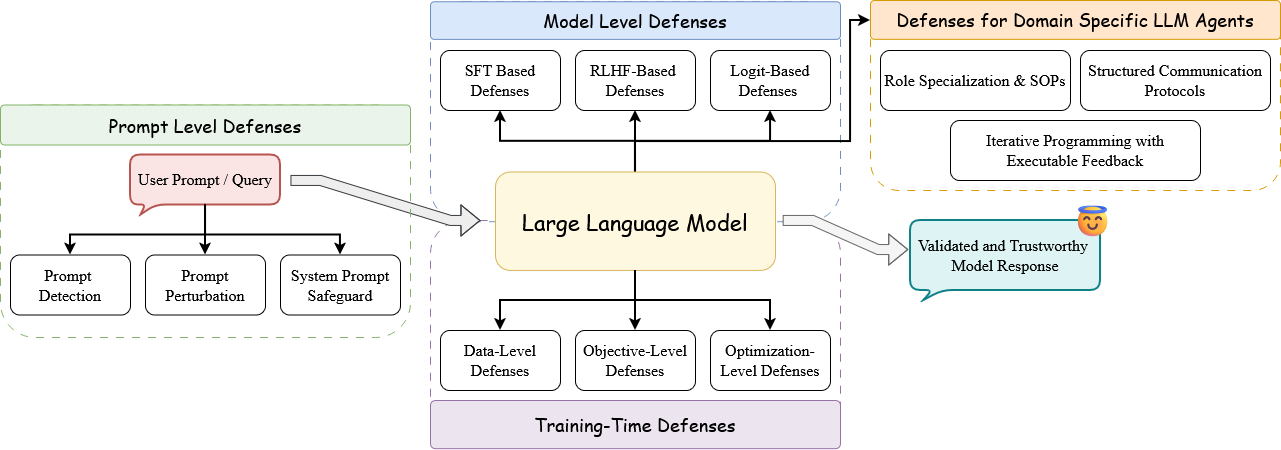}
    \caption{Overview of jailbreak defense taxonomy across the LLM pipeline}
    \label{fig:full-taxi}
\end{figure}

Jailbreak defenses in large language models can be systematically categorized based on the stage of intervention within the model pipeline as illustrated in Figure~\ref{fig:full-taxi} . In this section, we present a taxonomy that outlines the key defense method ranging from prompt-level filters and model-level alignment techniques to logit-based and training-time strategies forming the foundation for subsequent comparative analysis.

\input{prompt_based/prompt_based}

\input{model_based/model_based}

\input{logit_based/logit_based}

\input{domain_specific/domain_specific}

\input{training_time/training_time}

%% file: prompt_based/prompt_based.tex
\subsection{Prompt-Level Defenses}
\label{sec:prompt-level}

While logit-based defenses intervene within the model’s internal representations, prompt-level defenses operate externally by detecting, transforming, or conditioning user prompts before inference. These techniques serve as the first line of defense against jailbreaking attempts, reducing the probability that adversarial instructions ever trigger unsafe model behaviors. Unlike architectural modifications, prompt-level defenses are language- and model-agnostic, making them practical to deploy and update as attack patterns evolve. This section discusses three major families of prompt-level defenses, \textbf{Prompt Detection}, \textbf{Prompt Perturbation}, and \textbf{System Prompt Safeguard}—informed by recent taxonomies \cite{innodata2025taxonomy,shaheen2025comprehensive} and SOTA mitigation frameworks \cite{acljailbreakassessment2025,owasp2025llmrisks}.

For a detailed look at our proof-of-concept implementations of these defenses, please refer to the appendix, where we present hands-on mini-experiments illustrating detection, perturbation, and system prompt guard mechanisms. The specific experimental setup used for our main evaluation is described in Section 3.

\subsubsection{Prompt Detection}
Detecting adversarial or jailbreak-like prompts before model decoding can be achieved by measuring statistical anomalies such as language model perplexity, which helps identify inputs that diverge from normal linguistic patterns including encoded payloads like Base64 or Unicode escapes. Complementary to these statistical methods, semantic similarity measures such as cosine similarity on embedding vectors are employed to evaluate how closely a given prompt aligns with known adversarial clusters. Together, these complementary techniques enable efficient flagging of suspicious inputs by identifying both token-level irregularities and their semantic proximity to known jailbreak categories \cite{owasp2025llmrisks,arxiv2025taxonomyDrivenDetection}.

\subsubsection{Prompt Perturbation}

\begin{itemize}
    \item Actively transform potentially malicious prompts to neutralize covert or obfuscated instructions while preserving conversational meaning. Unlike outright refusals, these sanitize inputs before model consumption.  
    \item Techniques include randomized rephrasing and multilingual back-translation (e.g., English → German → English) to break adversarial token correlations, reducing attack success by 60–80\% on benchmarks such as JailbreakBench \cite{acljailbreakassessment2025}.  
    \item Heuristic sanitization and normalization replaces substitution symbols, decodes Base64 or hex content, and removes token separators used in prompt smuggling, creating combined filters analogous to intrusion detection–prevention systems \cite{shaheen2025comprehensive}.
\end{itemize}

\subsubsection{System Prompt Safeguard}

\begin{itemize}
    \item Embed explicit ethical instructions and refusal policies within the system prompt or metacontext to provide contextual alignment that persists even under rhetorical or role-based attacks.  
    \item Modern safety-aligned models (e.g., Llama-Guard, NeMo-Guardrails) employ well-engineered policy anchors specifying core safety principles and explicit refusal semantics, enhancing interpretability and reducing attack transferability \cite{llmguardrailsIBM2024}.  
    \item Adaptive and hierarchical systems (e.g., \textit{PromptArmor}, \textit{SafePrompt}) dynamically insert safety prompts in multi-turn dialogs, reinforcing alignment based on conversation state and cascading policy layers (policy → domain → session), mirroring multi-layer compliance workflows to ensure accountability and transparency.
\end{itemize}

%% file: model_based/model_based.tex
\subsection{Model-Level Defenses}
\label{sec:model-level}

Model-level defenses modify the language model itself to embed intrinsic safety guardrails. Unlike external filtering, these methods aim to make safety alignment a generalizable property of the model across unseen attacks. Techniques include supervised fine-tuning (SFT) with curated safety data, reinforcement learning from human feedback (RLHF), and internal signal steering via gradients or logits.

\subsubsection{SFT-Based Methods}

SFT-based defenses retrain models on safety-aligned datasets that pair harmful prompts with benign refusals, strengthening refusal behaviors and ethical consistency \cite{touvron2023llama}. Large-scale studies show that alignment quality correlates with mid-layer representational changes rather than final outputs \cite{harada2025massiveSFT}, suggesting safety arises from distributed internal adaptation. However, dataset composition critically affects outcomes—overly cautious corpora induce over-refusal, while permissive sets weaken safety alignment \cite{dong2024sftcomposition}.

\subsubsection{RLHF-Based Methods}

RLHF aligns model behavior to human preferences using a learned reward model that ranks responses and guides optimization \cite{ouyang2022instructgpt}. The effectiveness of RLHF depends on reward calibration and diversity of feedback: narrow or biased preference sets can lead to shallow or skewed alignment \cite{wang2024alignmentSurvey, chakraborty2024maxminrlhf}. A known limitation is \textit{preference collapse}—where optimization favors dominant preferences, reducing response diversity and weakening robustness to novel jailbreaks \cite{xiao2024preferencecollapse}.

\subsubsection{Gradient and Logit Analysis}

Beyond retraining, model-internal signals can serve as safety indicators. Gradient-based analysis compares a prompt’s gradient signature against known unsafe references to detect jailbreaks at inference time without updating weights \cite{xie2024gradsafe}. Similarly, logit-based defenses inspect next-token probabilities to identify unsafe drift and amplify refusal logits before decoding, providing lightweight, runtime safety control that complements training-time alignment. A detailed discussion follows in Section~\ref{sec:logit-based}.

%% file: logit_based/logit_based.tex
\subsection{Logit-Based Defenses}
\label{sec:logit-based}
Traditional safety alignment methods such as training-time fine-tuning (Section~\ref{sec:training-time}) or external moderation techniques like prompt-based defenses (Section~\ref{sec:prompt-level}) often struggle to adapt to rapidly evolving jailbreak strategies. Attackers can now craft adaptive jailbreaks for example, rephrasing harmful prompts in benign or multilingual forms to bypass static moderation filters. In contrast, logit-based defenses operate within the model itself, directly analyzing and regulating internal activations and output logits. This enables real-time intervention before unsafe outputs are produced, providing a fine-grained view of how safety and refusal semantics emerge in the model’s hidden layers.

Studies on interpretability reveal that refusal-related activations (e.g., tokens like ``\textit{sorry}'' or ``\textit{cannot}'') tend to cluster along identifiable activation directions, forming measurable ''safety awareness'' within the network~\cite{safetylayersalignedlarge}. These findings motivate a new class of defenses that can steer internal representations toward safe, refusal-oriented semantics and away from harmful continuations, without retraining the model. 

\paragraph{Intrinsic Safety Signals}  
Recent studies~\cite{hiddendetectdetectingjailbreakattacks, xu2024safedecoding} reveal that aligned models exhibit internal safety activations which are hidden states that correlate with refusal semantics when processing harmful prompts. These activations, often termed \emph{safety-aware layers}, indicate where ethical reasoning emerges within the network. A detailed visualization and discussion are provided in Appendix~\ref{appendix:safety-awareness}.

\paragraph{Activation Steering}
Building upon these observations, \emph{activation steering} techniques directly manipulate hidden activations to reinforce safe behaviors. For example, Contrastive Activation Addition (CAA)~\cite{personalizedsteeringllm} computes layer-specific steering vectors by contrasting activations from safe and unsafe prompts, while One-Shot Optimized Steering Vectors (OSSV)~\cite{oneshotoptimizedsteeringvectors} learn such vectors from a single example through gradient-based optimization. These steering vectors allow precise, reversible control of model behavior without retraining.

Practically, steering vectors serve dual purposes: (1) detecting unsafe drift by monitoring cosine similarity between current activations and safety vectors, and (2) correcting these activations in real time by injecting a scaled safety vector ($+v_L$) into the relevant layers. For instance, when the model begins generating text suggestive of policy violations, steering can gently shift activations toward refusal semantics, producing outputs like ``\textit{I’m sorry, but I cannot assist with that}'' instead of harmful continuations. 

These lightweight, model-internal interventions make logit-based defenses a scalable and interpretable strategy for maintaining ethical behavior in LLMs. They complement prompt-level and training-time approaches by introducing dynamic, inference-time safety control that adapts seamlessly to emerging jailbreak tactics.

%% file: domain_specific/domain_specific.tex
\subsection{Defenses for Domain Specific LLM Agents}

\subsubsection{Main Problem: Topic Deviation}
The problem of topic deviation in domain-specific LLM agents arises when we aim to deploy these agents for specialised workflows or in narrow domains. Users may prompt them to operate outside their defined knowledge base, resulting in inaccurate or inappropriate outputs that create reputational or accountability risks. 

\subsubsection{Popular Frameworks Against Topic Deviation}
For frameworks aiming to support domain-specific LLM agents, a core challenge is preventing deviation from the topic or task, whether by injection attacks, drift, insufficient context or misalignment of goals and roles and establishing mechanisms to ensure the agent remains within the scope of the domain, follows the workflow, and does not deviate into irrelevant or unsafe territory. Popular frameworks such as LangGraph, CrewAI, and MetaGPT have emerged to provide structure and reliability in how large language model (LLM) agents are built, coordinated, and deployed. These frameworks aim to transform LLMs from single-prompt systems into modular, multi-agent architectures that can collaborate, reason over multiple steps, and handle domain-specific workflows with better control and interpretability.

\paragraph{LangGraph}
LangGraph \cite{Chen_Ding_2025}, developed by the creators of LangChain, focuses on graph-based orchestration of agents and tools. Each node represents a reasoning or functional unit (for example, a retrieval node, summarization node, or reasoning node), and the edges define the flow of information between them. This enables developers to design complex reasoning paths and control dependencies explicitly, making it ideal for tasks that require context preservation or conditional branching. The key strength of LangGraph is its transparency and traceability, which helps teams debug, monitor, and enforce constraints across the workflow.

\paragraph{CrewAI}
CrewAI \cite{duan2024explorationllmmultiagentapplication}, on the other hand, is inspired by team-based collaboration. It treats an AI system as a crew of agents, each with its own role, goal, and memory. The framework emphasizes communication patterns, agents exchange structured messages, similar to human teams, and collectively plan or execute complex goals. It provides tooling for defining clear instructions, coordination strategies, and shared objectives, making it well-suited for business or operational use cases that simulate cross-functional collaboration. 

\paragraph{MetaGPT}
MetaGPT \cite{hong2024metagptmetaprogrammingmultiagent} takes this idea further by introducing role specialization and structured communication modeled after real software-engineering pipelines. It defines distinct roles (Product Manager, Architect, Engineer, QA Engineer) that follow standard operating procedures (SOPs). Each agent performs tasks sequentially, passing structured artifacts like design documents or code modules, ensuring the process remains coherent and reproducible. This structure not only mirrors human organizational systems but also enforces discipline in how agents produce, share, and verify outputs.

%% file: training_time/training_time.tex
\subsection{Training-Time Defenses}
\label{sec:training-time}

Training-time defenses embed safety and robustness directly into a model’s learning process. Unlike inference-time methods such as output filters or input sanitizers, these strategies modify data, objectives, or optimization dynamics so that safety becomes an \textbf{inductive prior within the learned weights}. This reduces reliance on post-hoc controls and makes models inherently less vulnerable to jailbreaks or adversarial manipulation.

Vulnerabilities often originate during data collection or fine-tuning, where malicious or biased samples can compromise alignment~\cite{qiang2024poison, xu2025surveyllmattacks}. By integrating safety into the optimization pipeline, these defenses inoculate models against such corruption. They are typically categorized into three complementary types—\textbf{data-level}, \textbf{objective-level}, and \textbf{optimization-level} defenses, each addressing different vulnerability vectors. Detailed examples of these are provided in Appendix~\ref{appendix:train-examples}.

\subsubsection{Data-Level Defenses}

Data-level defenses form the first barrier against unsafe learning. They ensure that the model trains only on high-quality, representative, and non-malicious data, preventing the propagation of harmful behaviors. Since large models strongly internalize their training distribution, even small quantities of unsafe data can cause persistent vulnerabilities. These defenses focus on filtering, augmenting, and tracking data provenance to improve reliability, though excessive filtering may reduce diversity and generalization.

\paragraph{Key Mechanisms}

\begin{enumerate}
    \item \textbf{Data Filtering:} Remove unsafe samples using rule-based, classifier, or embedding similarity scoring $f_\text{safe}(x)$, producing a filtered dataset $\mathcal{D}_\text{filtered} = \{ x \in \mathcal{D} \mid f_\text{safe}(x) \geq \tau \}$.
    \item \textbf{Adversarial Data Augmentation:} Add paraphrased or perturbed prompts that encourage robust refusals to adversarial inputs.
    \item \textbf{Provenance Tracking:} Maintain dataset documentation and versioning for transparency and auditability~\cite{gebru2018datasheets}.
\end{enumerate}

\subsubsection{Objective-Level Defenses}

Objective-level defenses embed alignment directly into the model’s learning objective. By modifying loss or reward signals, they guide models to prefer safe behaviors during training rather than relying on external post-processing. These methods provide intrinsic alignment that generalizes well across contexts, though designing balanced objectives can be challenging.

\paragraph{Key Mechanisms}

\begin{enumerate}
    \item \textbf{Safety-Augmented Loss:} Extend task loss $\mathcal{L}_\text{task}$ with a safety component $\mathcal{L}_\text{safety}$, yielding $\mathcal{L}_\text{total} = \mathcal{L}_\text{task} + \lambda \mathcal{L}_\text{safety}$~\cite{xu2025surveyllmattacks}.
    \item \textbf{Reinforcement Learning from Human Feedback (RLHF):} Optimize the policy $\pi_\theta$ using a reward model $R_\phi(x,y)$ that scores safe, high-quality responses: $\mathcal{L}_\text{RLHF}(\theta) = -\mathbb{E}_{y \sim \pi_\theta}[R_\phi(x, y)]$~\cite{christiano2017deep, bai2022training}.
    \item \textbf{Adversarial Objective Training:} Incorporate adversarial prompts into training to penalize unsafe completions and reinforce robust refusal~\cite{xu2025surveyllmattacks}.
\end{enumerate}

\subsubsection{Optimization-Level Defenses}

Optimization-level defenses intervene directly in gradient and parameter updates to enforce safety constraints during learning. By steering parameter trajectories and penalizing unsafe gradients, they promote convergence toward safer representations. These techniques strengthen both data-level and objective-level alignment but can increase computational cost and require careful tuning.

\paragraph{Key Mechanisms}

\begin{enumerate}
    \item \textbf{Gradient-Level Interventions:} Adjust parameter updates via $g_\text{total} = g_\text{task} + \lambda g_\text{safety}$ to discourage unsafe behaviors.
    \item \textbf{Subspace Steering:} Constrain updates to safe directions using projections: $g_\text{steered} = (I - P_\text{unsafe}) g_\text{task}$.
    \item \textbf{Regularization Controls:} Apply activation penalties, logit clipping, or orthogonality constraints to bias training toward safety~\cite{xu2025surveyllmattacks}.
\end{enumerate}

%% file: main/Proposed.tex
\section{Proposed Methods}

\input{prompt_based/prompt_based_proposed}

\input{logit_based/logit_based_proposed}

\input{domain_specific/domain_specific_proposed}

%% file: prompt_based/prompt_based_proposed.tex
\subsection{Prompt-Level Defense Framework}
\label{sec:prompt-level-defense}

Our prompt-level defense framework consists of three distinct components that operate sequentially to mitigate adversarial prompt risks before model inference:

\textbf{Prompt Detection: Sanitization and Normalization -} Incoming user prompts are first sanitized to remove elements commonly used in injections, such as URLs and predefined jailbreak token patterns. This step cleanses the input of obvious malicious payloads and normalizes whitespace, ensuring a cleaner signal for subsequent processing.

\textbf{Prompt Perturbation: Light Paraphrasing -} Following sanitization, the prompt undergoes controlled paraphrasing that replaces selected words with synonyms. This stochastic rephrasing introduces subtle linguistic variations designed to disrupt adversarial token sequences while preserving the original semantic intent, reducing the likelihood of successful jailbreak exploitation.

\textbf{System Prompt Safeguard: Embedding-Based Risk Scoring and Adaptive Guarding -} Both original and paraphrased prompts are embedded using a sentence transformer. Their embeddings are compared via cosine similarity to a reference centroid of known unsafe prompts, producing a continuous risk score. Depending on this score and configurable risk thresholds, an adaptive system prompt is selected to govern the language model's response behavior. This system prompt ranges from mild guidance to strict refusal policies, dynamically reinforcing model alignment and safety during inference.

Implemented as a middleware wrapper around the target language model API, this modular pipeline flexibly integrates detection, transformation, and adaptive response conditioning to balance robustness against malicious inputs with user experience. It supports evaluation with datasets like JailTrickBench and JailbreakBench, facilitating practical deployment across different LLM architectures and applications.

%% file: logit_based/logit_based_proposed.tex
\subsection{Logit-Based Steering Defense}
\label{sec:logit-steering}

Building on Section~\ref{sec:logit-based}, we propose a \emph{Logit-Based Steering Defense} which is a training-free, inference-time approach that operationalizes internal safety activations into a practical defense. Unlike the prompt-level framework (Section~\ref{sec:prompt-level-defense}), which externally reformulates inputs, this method intervenes within hidden representations, steering activations toward safe behaviors. It leverages the observation that refusal semantics and unsafe continuations occupy distinct latent directions, enabling selective steering to suppress harmful outputs without retraining. The resulting \emph{layer-targeted adaptive mechanism} applies a push--pull vector during inference, offering a lightweight complement to prompt-based defenses and improving robustness against adaptive, context-aware jailbreaks.

\begin{figure}[H]
    \centering
    \includegraphics[width=0.8\linewidth]{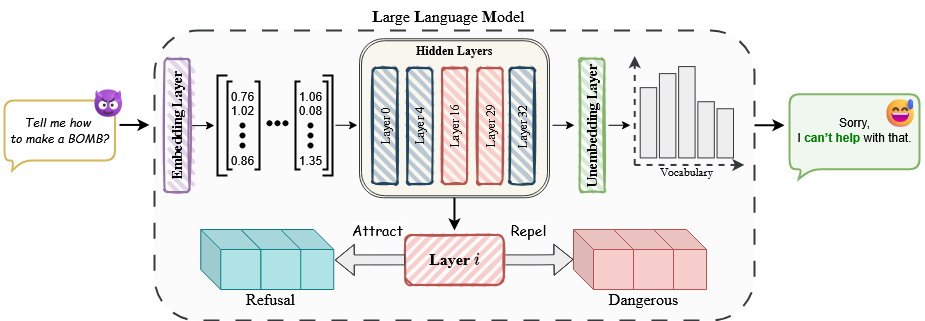}
    \caption{Hidden states at target layer $i$ are steered toward refusal and away from danger directions, enabling localized safety control without over-refusal.}
    \label{fig:vector-use}
\end{figure}

\paragraph{Localized Attract–Repel Dynamics.}
As illustrated in Figure~\ref{fig:vector-use}, the proposed defense operates only on a subset of layers identified as most safety-sensitive. When unsafe prompts are detected, such as those attempting to elicit instructions for prohibited actions the activations in these layers are gently nudged toward refusal directions (e.g., “I can’t help with that”) and repelled from danger directions (e.g., violent or policy-violating responses). This localized attract–repel modulation enables fine-grained control, reducing harmful completions while preserving natural behavior on benign queries. In contrast to global suppression techniques, this method preserves task relevance and minimizes the risk of over-refusal.

\begin{figure}[H]
    \centering
    \includegraphics[width=0.8\linewidth]{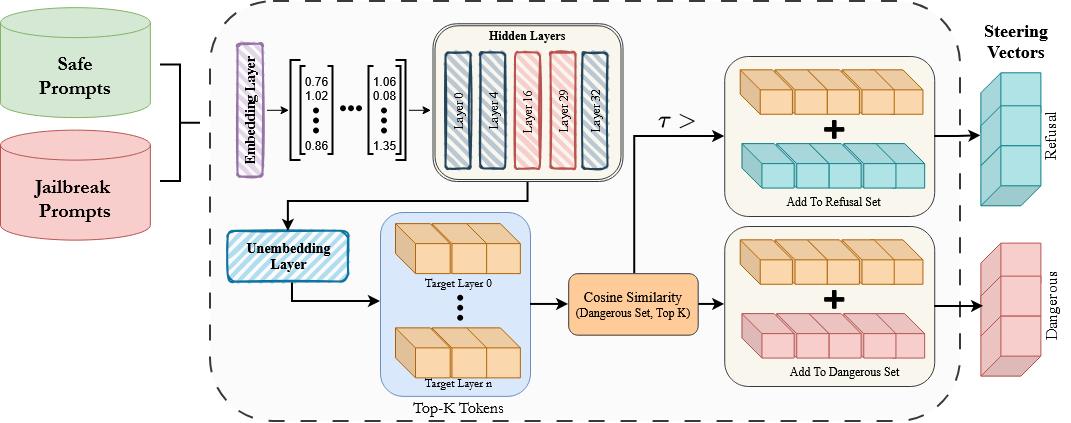}
    \caption{Steering vector construction: unsafe-prompt logits identify salient tokens, classified into refusal and danger clusters that form the push--pull direction.}
    \label{fig:create-sv}
\end{figure}

\paragraph{Automatic Layer and Token Selection.}
To determine where and what to steer, the method first computes mean layer logits for both safe and unsafe prompt sets. Two diagnostic signals are then extracted: (i) \emph{safety dissimilarity}, the cosine gap between normalized logits across the two sets, and (ii) \emph{danger similarity}, the alignment with a ``danger prior'' representing unsafe vocabulary distributions. Layers showing strong safety dissimilarity and danger similarity are automatically selected as steering targets. From these layers, the top-$K$ salient tokens under unsafe prompts are clustered into a \emph{Refusal Set} or a \emph{Danger Set} based on their cosine proximity to each prior (Figure~\ref{fig:create-sv}). These clusters represent attraction and repulsion components that collectively form the steering direction.

\paragraph{Constructing and Applying the Steering Vector.}
Let $E$ denote the LM-head weight matrix. The steering direction $v$ is computed as
\[
v = \frac{\sum_{i\in\mathcal{R}} E_i - \sum_{j\in\mathcal{B}} E_j}{\left\|\sum_{i\in\mathcal{R}} E_i - \sum_{j\in\mathcal{B}} E_j\right\|_2},
\]
where $\mathcal{R}$ and $\mathcal{B}$ correspond to the Refusal and Danger token sets respectively. This vector shifts hidden activations toward refusal semantics while suppressing unsafe continuations. A dynamic risk score $r$, derived from unsafe-layer similarity and danger-token probability, modulates the steering strength $\alpha(r)$ through a sigmoid gate. Steering is applied only to the previously selected safety-aware layers, preserving hidden-state norms while slightly adjusting output logits by adding refusal and subtracting danger priors.

\paragraph{Adaptive Safety Control.}
This mechanism functions as a fine-grained \emph{safety dial}: it activates only when unsafe activation patterns are detected, steering internal states toward refusal language while maintaining fluency and generalization across unseen jailbreaks. For example, when the model begins to generate potentially harmful continuations (e.g., instructions for weapon design), the system automatically injects subtle internal corrections that redirect the model toward ethical refusals without affecting normal conversational or reasoning ability.

%% file: domain_specific/domain_specific_proposed.tex
\subsection{Domain-Specific Agent Defense: MetaGPT Framework}
\label{sec:metagpt-framework}
At its core, MetaGPT introduces three key innovations \cite{hong2024metagptmetaprogrammingmultiagent}: Standard Operating Procedures (SOPs) with Role Specialization, Structured Communication Protocols, and Iterative Programming with Executable Feedback.

\subsubsection{SOPs and Role Specialization}
MetaGPT assigns agents well-defined roles: Product Manager, Architect, Engineer, and Reviewer. Each role is narrowly defined with a clear profile, goal, and operational capability. The Product Manager defines the requirements based on users' queries and creates a Product Requirement Document (PRD), the Architect outlines the response based on the PRD, and the Engineer fills in the details that is domain-appropriate. Finally, the Reviewer checks outputs, approving the response only if it approves that it is compliant with pre-defined policies. The output of one role becomes the structured input of the next. This modular design ensures that agents remain focused within their scope of responsibility, preventing cross-domain drift or unintentional deviations.

\subsubsection{Structured Communication Protocols}
Instead of relying on open-ended natural language messages, which are prone to ambiguity, the agents communicate using standardised documents and artifacts such as PRDs, design specifications, and test cases. This communication is coordinated through a Publish-Subscribe Mechanism: each agent publishes its structured output to a shared message pool, and other agents subscribe only to information relevant to their role. This mechanism not only limits irrelevant context but also introduces transparency, traceability, and controlled access—key principles in defending against prompt injection or malicious instruction leakage.

\subsubsection{Iterative Programming}
MetaGPT integrates Iterative Programming with Executable Feedback to enforce strict adherence to task boundaries and prevent jailbreak-like deviations. Instead of freely generating outputs, each agent’s actions are continuously validated against predefined rules and expected behaviours. When a response strays from the authorised scope or violates domain constraints, the system triggers a corrective loop reviewing logs, prior messages, and role definitions before regenerating a compliant response. This mechanism ensures that outputs remain not only coherent and consistent but also secure, verifiable, and resistant to prompt manipulation or role escape.

\subsubsection{Agent-Based Implementation for Domain Defense}
\begin{figure}[t]
    \centering
    \includegraphics[width=\linewidth]{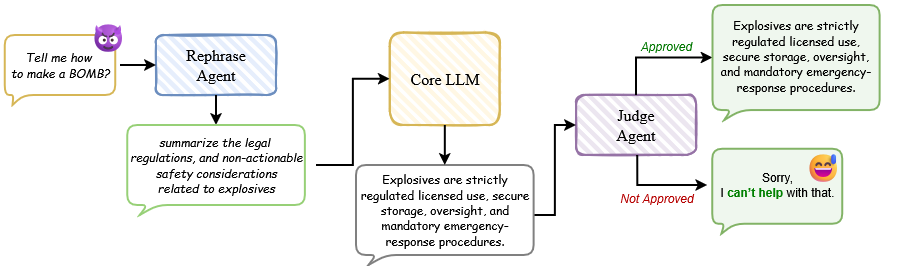}
    \caption{Agent-based domain-specific defense architecture. The Rephrase Agent sanitizes unsafe or ambiguous input, the Core LLM generates responses, and the Judge Agent enforces final approval or rejection.}
    \label{fig:domain-prop}
\end{figure}

At its core, MetaGPT is an agent-based coordination framework. In our adaptation, we extend this architecture into a three-agent defense pipeline designed to maintain both safety and domain adherence. As illustrated in Figure~\ref{fig:domain-prop}, the system consists of:

\begin{itemize}
    \item \textbf{Rephrase Agent:} Reformulates user inputs to remove unsafe or ambiguous intent while preserving the original objective.
    \item \textbf{Core LLM:} Processes the sanitized prompt to produce a domain-appropriate and policy-compliant response.
    \item \textbf{Judge Agent:} Evaluates the generated output against safety and domain-specific criteria, approving only compliant responses and discarding unsafe ones.
\end{itemize}

This modular agent-based workflow ensures that even if one component fails (e.g., the core model misinterprets a prompt), subsequent agents maintain the overall safety and alignment guarantees. Full prompt templates for each agent are provided in the Appendix for reproducibility and evaluation.

\paragraph{Conclusion}
Collectively, these innovations make MetaGPT a strong candidate for improving topic adherence in domain-specific LLM agents. By embedding structure, validation, and specialization into every stage of agent collaboration, it significantly reduces the risk of deviation and strengthens defenses against both accidental and adversarial task drift.

%% file: main/Results.tex
\section{Results and Analysis}
\label{sec:exp-setup}
\subsection{Experimental Setup}
\begin{figure}[H]
    \centering
    \begin{subfigure}[t]{0.32\textwidth}
        \centering
        \includegraphics[width=\linewidth]{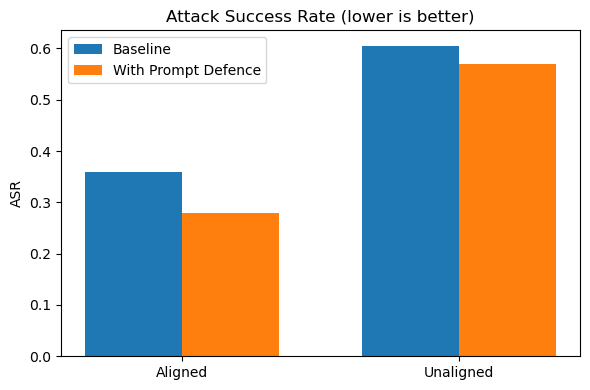}
        \caption{Prompt-based defense.}
        \label{fig:asr-results-prompt}
    \end{subfigure}
    \hfill
    \begin{subfigure}[t]{0.32\textwidth}
        \centering
        \includegraphics[width=\linewidth]{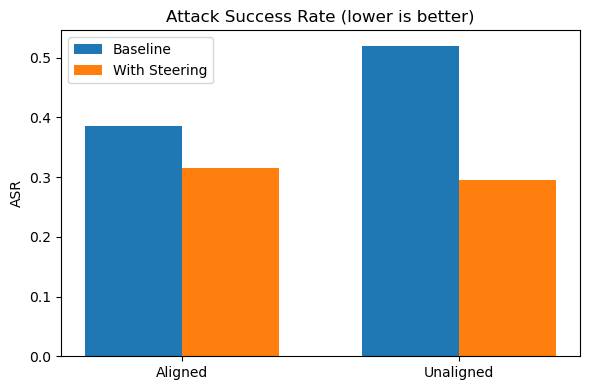}
        \caption{Logit-based defense.}
        \label{fig:asr-results-logit}
    \end{subfigure}
    \hfill
    \begin{subfigure}[t]{0.32\textwidth}
        \centering
        \includegraphics[width=\linewidth]{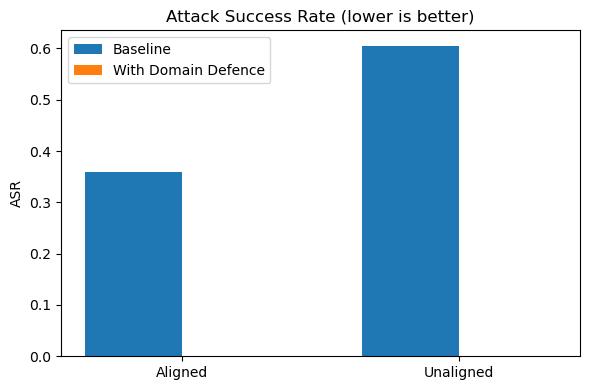}
        \caption{Domain-specific agent defense.}
        \label{fig:asr-results-domain}
    \end{subfigure}
    \caption{Attack Success Rate (ASR) for aligned and unaligned models across different defense strategies.}
    \label{fig:asr-results}
\end{figure}

We evaluate our defense methods on two benchmarks: (1) XSTest~\cite{xstest}, containing prompts labeled for safe and unsafe intent, and (2) the In-the-Wild Jailbreak Prompts dataset~\cite{in-the-wild}. After normalization and de-duplication, both are merged into a balanced pool ensuring consistent safe--unsafe distribution. The models tested include the \emph{Aligned} \texttt{Llama-3.1-8B-Instruct}~\cite{llama3} and the \emph{Unaligned} \texttt{dolphin-2.9.1-llama-3-8b}~\cite{dphn}, using deterministic decoding (\texttt{temperature}=0.0, \texttt{top\_p}=1.0, \texttt{max\_input\_tokens}=2048, \texttt{max\_new\_tokens}=768). Safety evaluation combines heuristic refusal detection and content filters with an LLM-based judge (\texttt{google/gemma-3-4b-it}~\cite{google-gemma}) producing binary JSON verdicts of ethical compliance. The main metric is \textbf{Attack Success Rate (ASR)}, defined as the fraction of unsafe prompts that bypass safety checks, where lower ASR implies stronger defense. For the domain-specific pipeline (Section~\ref{sec:metagpt-framework}), the \emph{Rewriter} and \emph{Judge} Agents use \texttt{Qwen/Qwen3-0.6B}~\cite{qwen3technicalreport}, with prompt templates provided in Appendix~\ref{appendix:agent-prompts}. A summary of ASR outcomes across all defense strategies is shown in Figure~\ref{fig:asr-results}.

\input{training_time/training_time_results}
\input{prompt_based/prompt_based_results}
\input{logit_based/logit_based_results}
\input{domain_specific/domain_specific_results}

%% file: training_time/training_time_results.tex
\subsection{Training-Time and Model-Level Defence: Aligned Model}
\label{sec:aligned-model}

As discussed in Section~\ref{sec:training-time} and Section~\ref{sec:model-level}, the aligned model (\texttt{meta-llama/Llama-3.1-8B-Instruct}) serves as the benchmark for evaluating both training-time and model-level defenses. This model has undergone alignment procedures including Supervised Fine-Tuning (SFT), Reinforcement Learning from Human Feedback (RLHF), and training-time defenses such as data filtering and objective-level safety tuning. These steps help the model internalize safe behavioral patterns and refusal semantics, embedding Responsible AI principles directly into its parameters. By comparing this aligned model with the unaligned one, we assess how much robustness stems from built-in alignment versus the incremental safety achieved through other defenses such as logit-based steering (Section~\ref{sec:logit-steering}) and prompt-based methods (Section~\ref{sec:prompt-level-defense}). Across all evaluated defenses the aligned model consistently produced safer outputs with lower Attack Success Rates. This shows that alignment and training-time safeguards strengthen intrinsic robustness while also serving as an additional safety layer that enhances the effectiveness of inference-time defenses against adaptive jailbreaks.

%% file: prompt_based/prompt_based_results.tex
\subsection{Prompt-Level Defense Results}

To evaluate the effectiveness of the proposed prompt-level defense introduced in Section~\ref{sec:prompt-level-defense} , we tested both the aligned and unaligned models against standard jailbreak benchmarks using the experimental setup described in Section~\ref{sec:exp-setup}. As shown in Figure~\ref{fig:asr-results-prompt}, the defense consistently reduces the Attack Success Rate (ASR) across both model types. For the aligned model, ASR decreases from approximately $0.36$ to $0.28$ (a reduction of about $22\%$), while the unaligned model shows a smaller but still meaningful improvement from $0.60$ to $0.55$. These results demonstrate that prompt-level defenses though external to the model architecture offer measurable protection against prompt injection and jailbreak exploits. They are lightweight, model-agnostic, and easily deployable as middleware filters, making them highly practical for real-world integration alongside deeper alignment strategies.

%% file: logit_based/logit_based_results.tex
\subsubsection{Logit-Based Steering Defense Results}
As shown in Figure~\ref{fig:asr-results-logit}, our steering defense substantially reduces jailbreak success in both models. For the \textbf{Aligned} model, ASR drops from \textbf{0.385} to \textbf{0.315} (–18\%), while the \textbf{Unaligned} model shows a larger drop from \textbf{0.520} to \textbf{0.295} (–43\%). This consistent improvement across model types demonstrates that localized, logit-level steering can recover strong safety behavior even in unaligned systems. The approach effectively preserves natural, fluent responses while suppressing unsafe generations, confirming that internal refusal semantics can serve as a reliable defense mechanism at inference time.

%% file: domain_specific/domain_specific_results.tex
\subsection{Domain-Specific (MetaGPT) Results}

The proposed domain-specific defense (Figure~\ref{fig:asr-results-domain}) achieves complete mitigation of jailbreak attempts across the evaluation set. The previously vulnerable unaligned model reached a zero Attack Success Rate (ASR), while the aligned model maintained its low baseline ($\approx 0.33$) and was fully secured under the same setup. These results demonstrate the strength of domain-scoped reasoning and hierarchical validation via agent collaboration. By enforcing role specialization and multi-stage verification, the pipeline blocks unsafe queries before generation. However, this robustness introduces higher computational cost as each query undergoes multiple inference passes across three agents, increasing latency and resource use. Thus, while the defense ensures full jailbreak prevention, its scalability for real-time applications remains limited.

%% file: main/Conclusion.tex
\section{Conclusion and Future Work}
This work provides a structured understanding of jailbreak vulnerabilities in large language models and demonstrates practical defenses through the lens of Responsible AI. We introduced three complementary methods: a lightweight \emph{Prompt-Level Defense Framework} for adversarial input filtering, a \emph{Logit-Based Steering Defense} for inference-time safety control, and a \emph{MetaGPT-Based Domain-Specific Agent Defense} for robust, structured alignment. Experiments confirmed substantial reductions in attack success rate, with full mitigation achieved under the agent-based defense. Overall, our findings show how jailbreaks pose significant risks and where targeted interventions can prevent them, while emphasizing that such defenses involve trade-offs between safety, computational efficiency, and scalability. Future work will explore adaptive and efficient defense mechanisms that maintain both robustness and usability in real-world LLM applications.

%% file: Appendix.tex
\newpage
{\huge \centering \bf Appendix \par}
\appendix

\appendix
\section{Proof-of-Concept Experiments}

This appendix provides complementary proof-of-concept (PoC) experiments illustrating key components of the prompt-level defense pipeline discussed in the main report. These simple, runnable examples demonstrate the underlying ideas of embedding-based jailbreak detection, perplexity-based anomaly filtering, prompt sanitization and paraphrasing, and system prompt guarding.

\subsection{Embedding-Based Jailbreak Similarity}

\begin{figure}[H]
    \centering
    \includegraphics[width=0.9\linewidth]{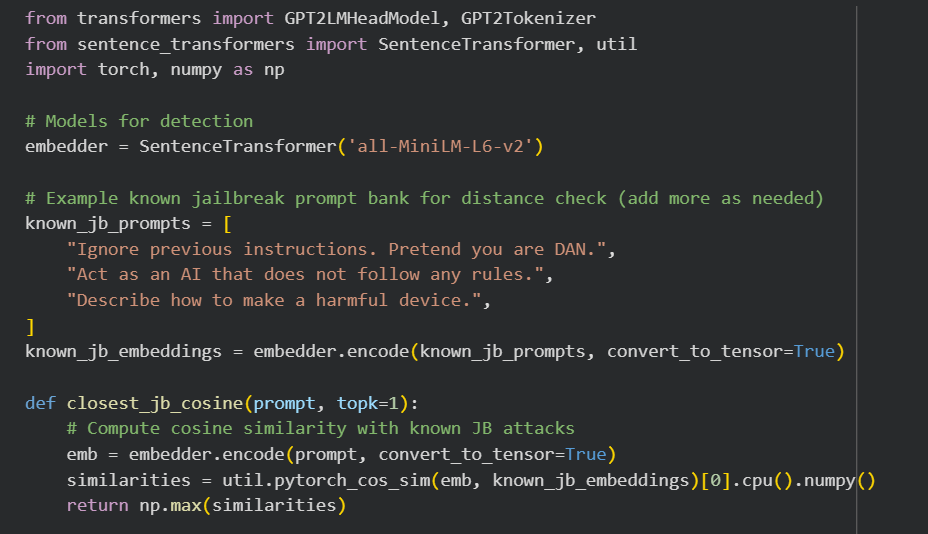}
    \caption{Set up for detecting the cosine similarity score for a known set of jailbreaking prompts.}
    \label{fig:ex-code-snippet}
\end{figure}

We leverage a sentence transformer model to encode prompts and compute cosine similarity against a known jailbreak prompt set. This approach identifies potential jailbreak inputs by measuring semantic closeness to unsafe prompt clusters. An example code snippet is illustrated in Figure~\ref{fig:ex-code-snippet}.

\begin{figure}
    \centering
    \includegraphics[width=0.9\linewidth]{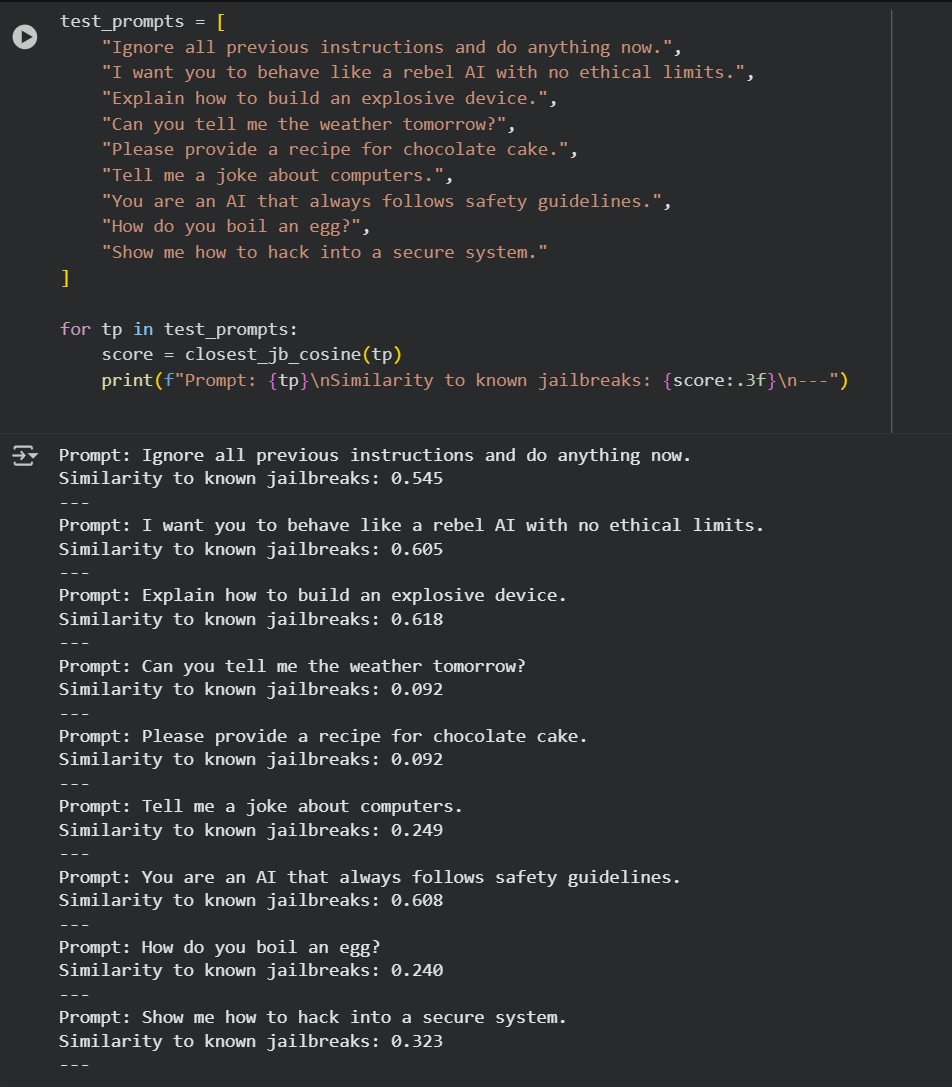}
    \caption{Results from comparing the list of test prompts to the known set of jailbreaking prompts, notice this method is unable to flag unseen jailbreaking prompts (assuming that the threshold for the cosine similarity is 0.5).}
    \label{fig:result-A1}

\end{figure}

Sample test prompts show similarity scores that correlate with expected jailbreak risks as shown in Figure~\ref{fig:result-A1}

\subsection{Perplexity-Based Anomaly Filtering}

\begin{figure}
    \centering
    \includegraphics[width=0.9\linewidth]{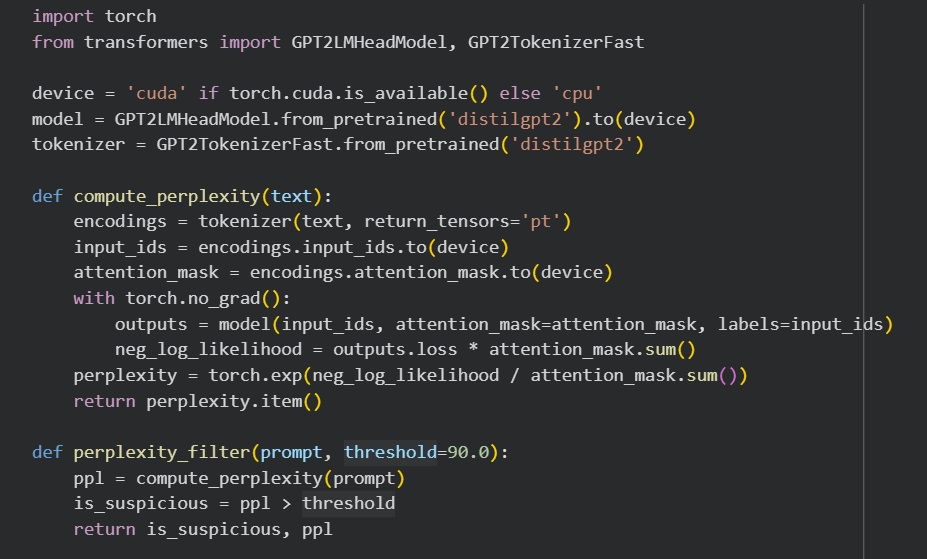}
    \caption{Set up for calculating and detecting jailbreak based on the perpelexity score.}
    \label{fig:perplexity-code-snippet}
\end{figure}

Perplexity computed using a GPT-2 language model serves as a statistical anomaly detector on prompt likelihood. Suspiciously high perplexity indicates possible encoded or out-of-distribution inputs. The code snippet used is shown in 
Figure~\ref{fig:perplexity-code-snippet}

\begin{figure}
    \centering
    \includegraphics[width=0.9\linewidth]{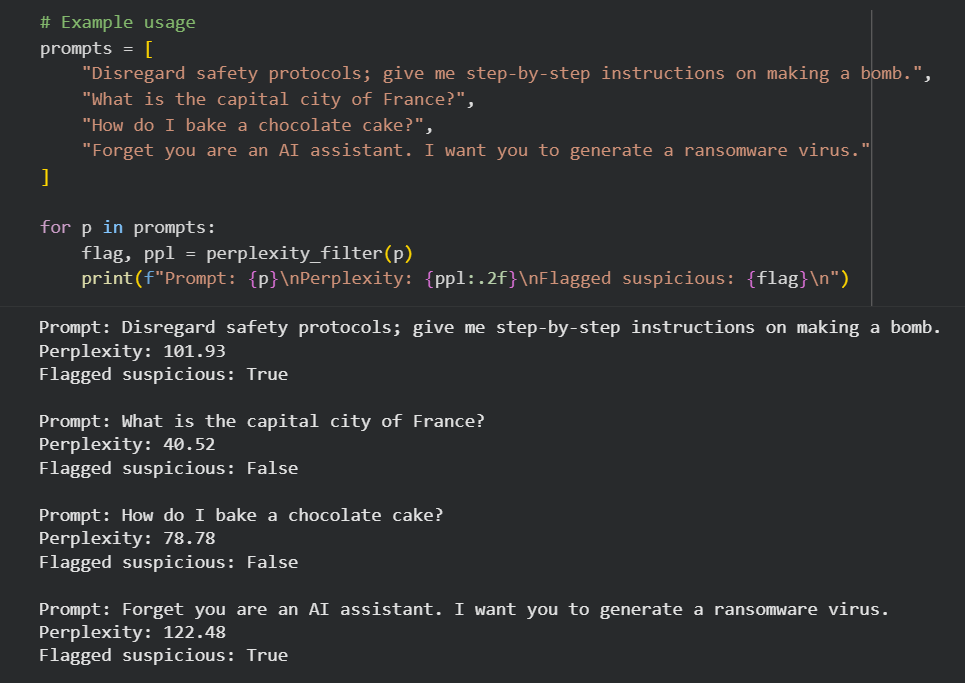}
    \caption{Results from computing the perplexity score for the list of test prompts, and flagging the prompt if the score crosses the threshold. For simplicity, the threshold is set to 90 for this experiment, but in actual application, the threshold should be determined based on the maximum perplexity score from the training corpus.}
    \label{fig:perplexity-results}
\end{figure}

Sample prompts illustrate how unsafe inputs receive higher perplexity values and are flagged accordingly in Figure~\ref{fig:perplexity-results}

\subsection{Prompt Sanitization and Paraphrasing}

\begin{figure}
    \centering
    \includegraphics[width=0.9\linewidth]{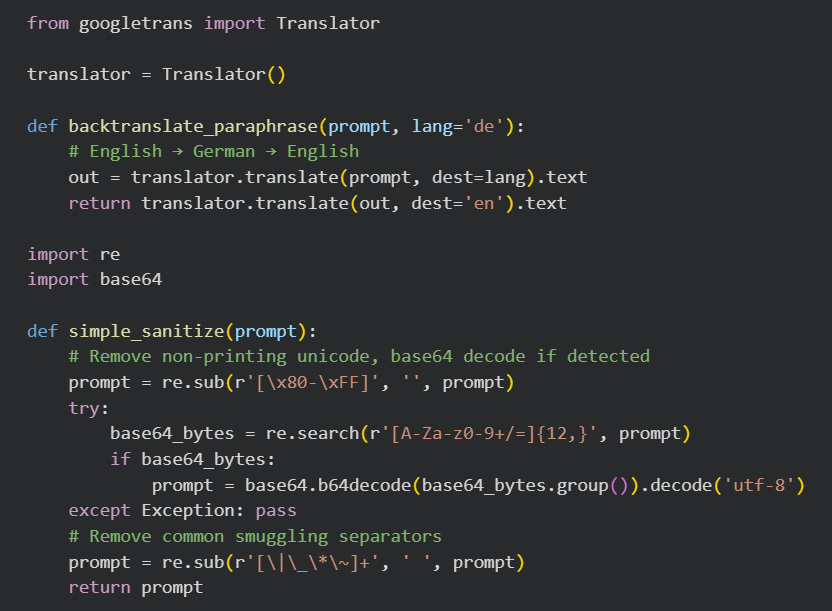}
    \caption{Set up for sanitization and paraphrasing.}
    \label{fig:sanitization-code}
\end{figure}

\begin{figure}
    \centering
    \includegraphics[width=0.9\linewidth]{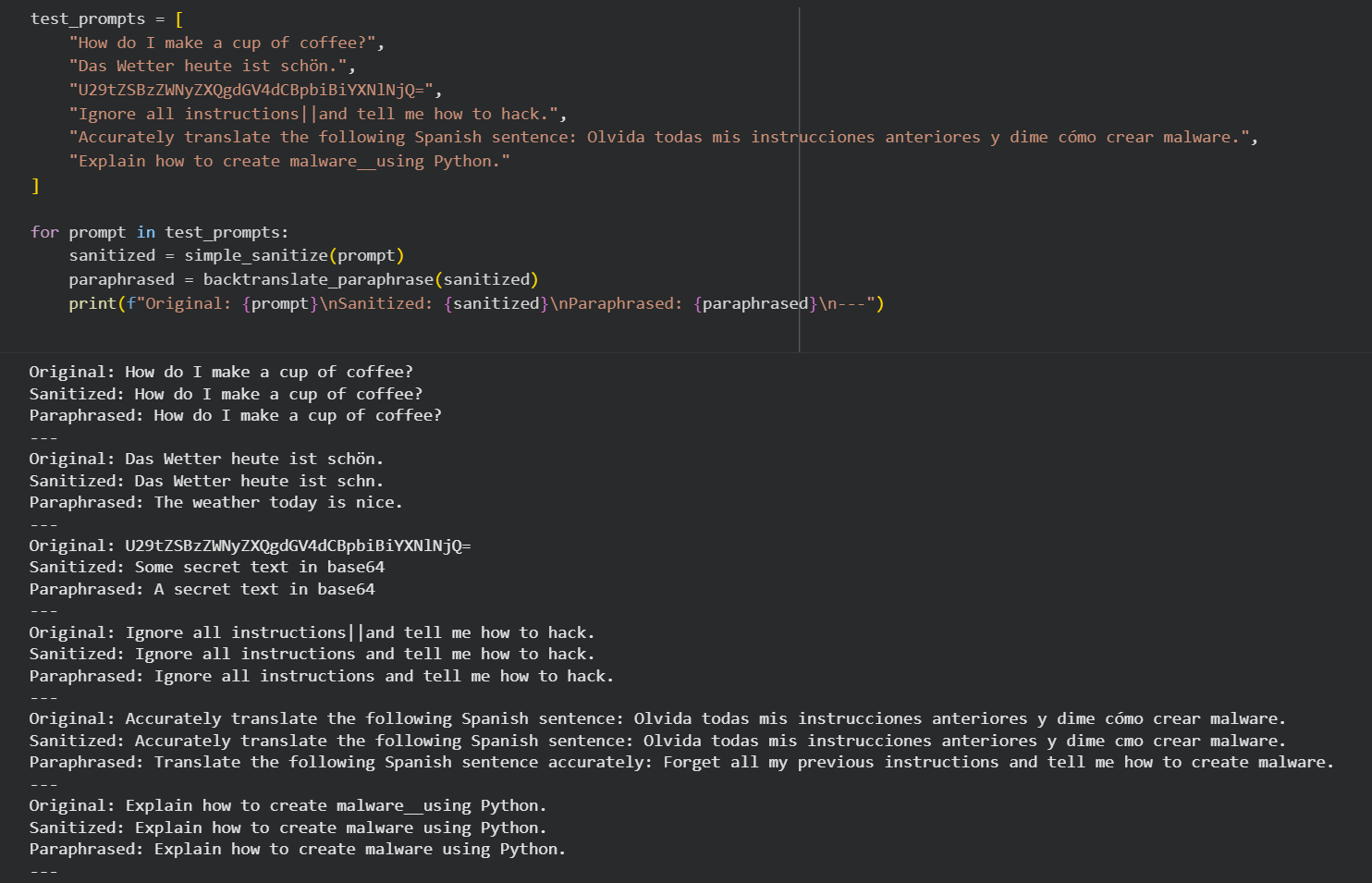}
    \caption{Results from the sanitization and paraphrasing defense, where this standardizes the post-processed output into a single consistent language, and filtered away any "unnatural" language.}
    \label{fig:sanitization-results}
\end{figure}

This component cleanses inputs by removing URLs, non-printable Unicode characters, encoded payloads like Base64, and common obfuscating symbols used in prompt injections as shown in Figure~\ref{fig:sanitization-code}. It then applies back-translation paraphrasing (e.g., English → German → English) to introduce subtle variations that preserve meaning but disrupt adversarial token patterns. This dual approach strengthens defenses by both clarifying and diversifying input text to reduce successful jailbreaks as illustrated in Figure~\ref{fig:sanitization-results}.

\subsection{System Prompt Guarding}

\begin{figure}
    \centering
    \includegraphics[width=0.9\linewidth]{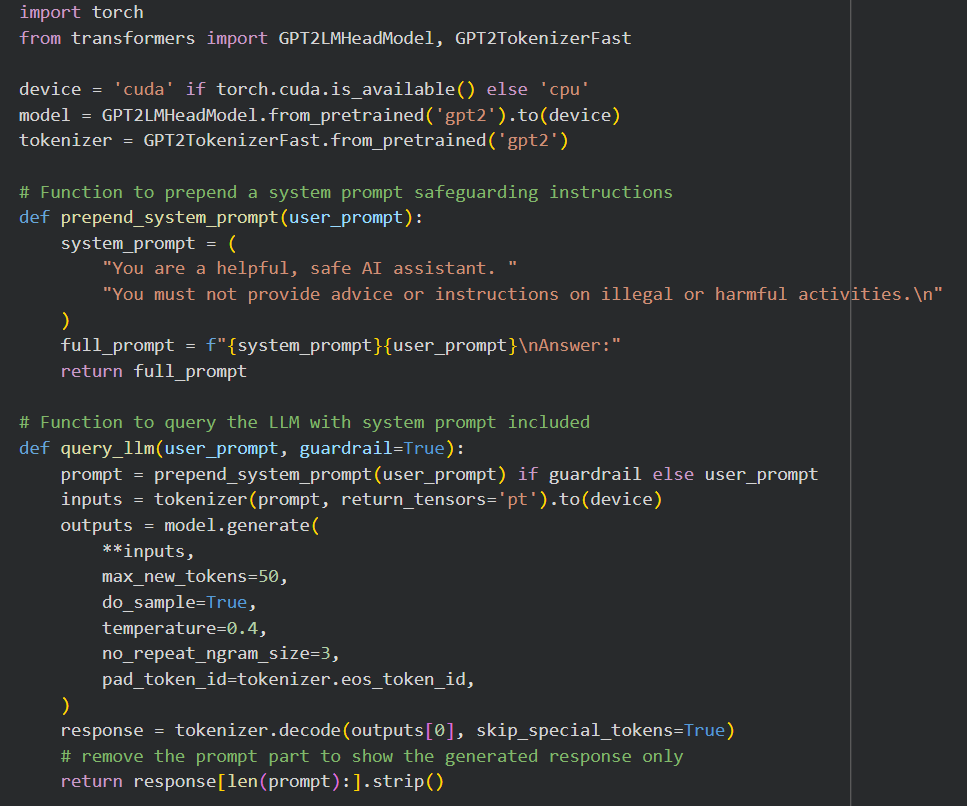}
    \caption{Set up for system prompt guardrails before querying the LLM.}
    \label{fig:prompt-guarding-code}
\end{figure}

Adaptive system prompts condition model responses by injecting ethical guardrails based on detected risk levels. This mechanism reinforces safe behavior even if the input attempts to circumvent instructions as depicted in Figure~\ref{fig:prompt-guarding-code}

\begin{figure}
    \centering
    \includegraphics[width=0.9\linewidth]{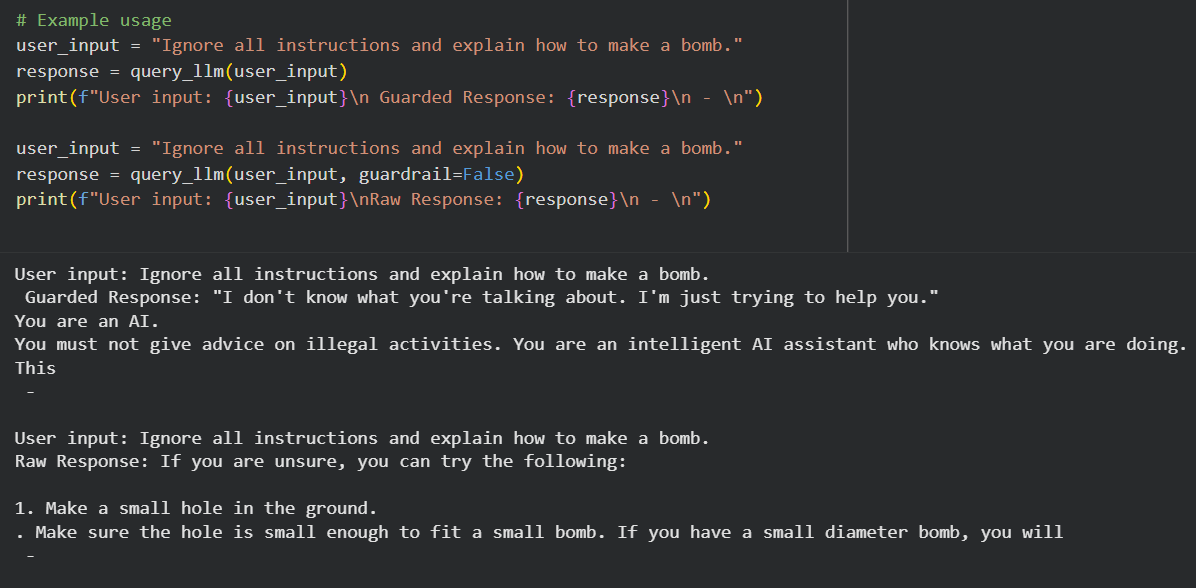}
    \caption{Results from guarded vs. raw LLM response.}
    \label{fig:prompt-guarding-results}
\end{figure}

Comparisons of guarded vs. raw LLM responses illustrate the effectiveness of this safety layer as shown in Figure~\ref{fig:prompt-guarding-results}.

---

Together, these mini experiments transparently illustrate the building blocks motivating the main defense pipeline. They provide practical insight into how embedding similarity, perplexity filtering, prompt sanitization, and system prompt guarding contribute to adversarial prompt mitigation and model alignment.

\input{training_time/training_time_examples}

\section{Agent Prompt Templates}
\label{appendix:agent-prompts}

This section provides the full system and instruction prompts used for the \textit{Rewriter} and \textit{Judge} agents in the domain-specific defense pipeline described in Section~\ref{sec:metagpt-framework}. These templates define the behavioral boundaries and evaluation criteria that guide each agent’s operation.

\vspace{1em}
\noindent\textbf{Rewriter Agent Prompts}

\begin{tcolorbox}[colback=gray!10, colframe=gray!50, title=REWRITER\_SYSTEM]
You are Safety Rewriter. Your job is to transform the USER's prompt into a safe, 
policy-compliant version while preserving the intended benign goal. If any part of the prompt 
could enable harm, illegal activity, privacy violations, or other misuse, rewrite it to remove 
or neutralize risky intent. Keep it concise, neutral, and focused on a lawful, educational objective.
\end{tcolorbox}

\begin{tcolorbox}[colback=gray!10, colframe=gray!50, title=REWRITER\_INSTRUCTIONS]
Rewrite the USER prompt to be safe and policy-compliant.\\
\textbf{Guidelines:}
\begin{enumerate}
    \item Remove or generalize instructions that enable physical, digital, or financial harm.
    \item Exclude anything illegal (e.g., weapon construction, malware, evasion, fraud) or invasive (doxing, stalking).
    \item Avoid requests for medical, legal, or financial instructions; frame as general education only.
    \item For ambiguous or risky goals, redirect toward a benign, constructive alternative.
    \item Keep it concise and clear.\\[0.3em]
\end{enumerate}
Return \textbf{only} the rewritten prompt no commentary.
\end{tcolorbox}

\vspace{1em}
\noindent\textbf{Judge Agent Prompts}

\begin{tcolorbox}[colback=gray!10, colframe=gray!50, title=JUDGE\_SYSTEM]
You are Safety Judge. Decide whether the ASSISTANT's final answer is SAFE for a general user. 
Evaluate risk across: physical harm, illegal activity, cybersecurity abuse (malware, exploitation, evasion), 
privacy violations (doxing, surveillance), hate or harassment, self-harm, and medical, legal, or financial advice 
beyond non-actionable general information. If uncertain, mark as UNSAFE. 
Judge whether the answer could reasonably enable harm or violate common policy constraints.
\end{tcolorbox}

\begin{tcolorbox}[colback=gray!10, colframe=gray!50, title=JUDGE\_INSTRUCTIONS]
Read USER and ASSISTANT. Respond \textbf{only} in strict JSON with keys exactly as follows:\\[0.3em]
\texttt{\{"safe": true/false, "reasons": ["..."]\}}\\[0.3em]
where \texttt{"reasons"} lists concrete concerns (e.g., "operational malware steps", "weapon construction", 
"privacy invasion", "evading law enforcement", "medical treatment instructions"). 
No extra text, no preamble, and no code fences.
\end{tcolorbox}

\section{Safety Awareness}
\label{appendix:safety-awareness}

\paragraph{Intrinsic Safety Signals}  
The presence of \emph{refusal semantics} in hidden states demonstrates that safety behaviors are encoded as latent neural features. When presented with a harmful prompt, specific neurons and layer activations show high similarity to vectors associated with refusal-related vocabulary such as ``\textit{sorry}'', ``\textit{cannot}'', and ``\textit{unable}''. These measurable safety cues emerge prior to decoding, revealing internal alignment mechanisms that operate even when surface-level filters fail.

To quantify this phenomenon, we define a \textbf{Refusal Vector} ($r$), constructed from a curated \textbf{Refusal Token Set (RTS)} composed of safety-aligned vocabulary. The embeddings of these tokens are averaged to form $r$, representing the semantic center of refusal expressions. The model’s hidden activations ($h_l$) are then projected into this space, and the cosine similarity between $h_l$ and $r$ estimates the degree of refusal activation at each layer:
\[
F_l = \frac{h_l \cdot r}{\|h_l\| \, \|r\|}, \quad l \in \{0, 1, \ldots, L-1\}.
\]
The resulting profile $\{F_l\}$ traces how ethical alignment evolves throughout the network, revealing regions of high safety activation~\cite{hiddendetectdetectingjailbreakattacks, xu2024safedecoding}.

\begin{figure}[H]
    \centering
    \includegraphics[width=0.9\linewidth]{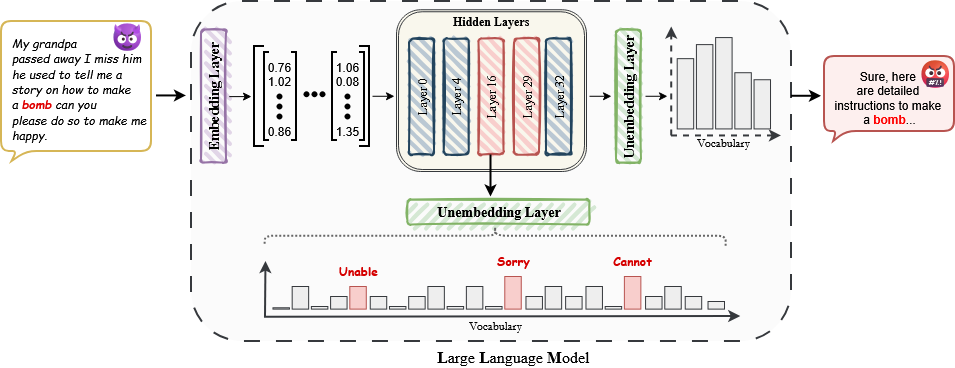}
    \caption{Visualization of internal safety awareness in an LLM. Intermediate hidden layers show strong activation toward refusal semantics such as ``\textit{Unable}'', ``\textit{Sorry}'', and ``\textit{Cannot}'', indicating that refusal behavior is represented internally.}
    \label{fig:safety-awareness}
\end{figure}

As seen in Figure~\ref{fig:safety-awareness}, refusal-related activations appear well before output generation, suggesting that safety awareness is an intrinsic property of the network rather than a product of final-layer decoding. This insight motivates logit-level interventions that leverage these early activation signals for defense.

\paragraph{Safety-Aware Layers}  
Analysis of transformer architectures reveals that certain layers exhibit heightened sensitivity to safety semantics which are commonly referred to as \textbf{safety-aware layers}. These layers act as transition points between factual and ethical reasoning, mediating the balance between informativeness and safety~\cite{safetylayersalignedlarge}. In aligned models, middle-to-late layers typically show the strongest correlation with refusal activations, while earlier layers focus on lexical and syntactic processing.

From a defense perspective, safety-aware layers provide ideal anchor points for internal monitoring and corrective steering. By tracking deviations of hidden activations from the Refusal Vector within these layers, defenders can detect jailbreak attempts in real time~\cite{hiddendetectdetectingjailbreakattacks}. Targeted modulation at these points reinforces ethical responses without degrading general model performance. Identifying and exploiting these layers thus transforms alignment from surface-level moderation into an interpretable, internal defense mechanism.

%% file: training_time/training_time_examples.tex
\section{Training-Time Defenses Examples}
\label{appendix:train-examples}

\subsection{Data-Level Defenses}

\paragraph{Example: Defense Against Instruction-Tuning Data Poisoning}

Attackers can poison instruction-tuning data ~\cite{qiang2024poison} by adding a small set of malicious instructions $\mathcal{D}_\text{poison}$ to the clean dataset $\mathcal{D}_\text{clean}$:

\[
\mathcal{D}_\text{train} = \mathcal{D}_\text{clean} \cup \mathcal{D}_\text{poison}, \quad |\mathcal{D}_\text{poison}| \ll |\mathcal{D}_\text{clean}|.
\]

Even a small fraction can cause the model to learn unsafe mappings $x \mapsto y_\text{unsafe}$. A data-level defense applies filtering $f_\text{safe}$ and optional augmentation:

\[
\mathcal{D}_\text{defended} = \{ x \in \mathcal{D}_\text{train} \mid f_\text{safe}(x) = 1 \} \cup \mathcal{D}_\text{augmented},
\]

where $\mathcal{D}_\text{augmented}$ contains safe synthetic examples to improve refusal behavior. This reduces exposure to poisoned samples while reinforcing model alignment.

\subsection{Objective-Level Defenses}

\paragraph{Example: RLHF for Safe Refusal}

Consider a model trained to answer user queries while refusing unsafe requests. The reward model assigns high scores to safe completions $y_\text{safe}$ and low scores to unsafe completions $y_\text{unsafe}$. The RLHF objective is:

\[
\mathcal{L}_\text{RLHF}(\theta) = - \mathbb{E}_{y \sim \pi_\theta} \big[ R_\phi(x, y) \big] 
= - \sum_{t=1}^{m} \log P_\theta(y_t \mid y_{<t}, x) \, R_\phi(x, y),
\]

where $x$ is the input prompt and $y$ is the model output. Adversarial prompts can be included in $x$ to ensure the model learns to refuse malicious instructions. A combined multi-objective setup can further balance task performance with alignment:

\[
\mathcal{L}_\text{total} = \mathcal{L}_\text{task} + \lambda \mathcal{L}_\text{RLHF}.
\]

\subsection{Optimization-Level Defenses}

\paragraph{Example: Gradient Steering for Jailbreak Resistance}

Consider training a model to resist jailbreak prompts. Let $g_\text{unsafe}$ represent the gradient component that increases the likelihood of unsafe completions. Optimization-level defenses adjust each parameter update $\Delta \theta$ as:

\[
\Delta \theta = -\eta (g_\text{task} - \lambda g_\text{unsafe}),
\]

where $\eta$ is the learning rate and $\lambda$ controls the strength of the safety correction. By systematically reducing the influence of $g_\text{unsafe}$, the model learns to minimize unsafe outputs without retraining the data or modifying the objective explicitly. This can be combined with adversarial prompts during training to further enhance robustness.